\documentstyle[12pt,a4,epsf]{article}

\begin{document}

\baselineskip=17pt plus 0.2pt minus 0.1pt

\makeatletter
\@addtoreset{equation}{section}
\renewcommand{\theequation}{\thesection.\arabic{equation}}
\renewcommand{\thefootnote}{\fnsymbol{footnote}}
\def\ast{a_*}
\def\ust{u_*}
\def\ae{a_{\rm eff}}
\def\phieff{\phi_{\rm eff}}
\def\calM{{\cal M}}
\def\calO{{\cal O}}
\def\calV{{\cal V}}
\def\calD{{\cal D}}
\def\p{{\partial}}
\def\nn{{\nonumber}}
\newcommand{\bea}{\begin{eqnarray}}
\newcommand{\eea}{\end{eqnarray}}
\begin{titlepage}
\title{
\hfill\parbox{4cm}
{\normalsize KUNS-1695\\{\tt hep-th/0011002}}\\
\vspace{1cm}
Descent Relation of Tachyon Condensation\\
from Boundary String Field Theory
}
\author{
Sanefumi {\sc Moriyama}${}^1$
\thanks{{\tt moriyama@gauge.scphys.kyoto-u.ac.jp}}
\quad and \quad
Shin {\sc Nakamura}${}^{1,~2}$
\thanks{{\tt nakamura@gauge.scphys.kyoto-u.ac.jp}}
\\[15pt]
${}^1$
{\it Department of Physics, Kyoto University, Kyoto 606-8502, Japan}
\\[10pt]
${}^2$
{\it The Graduate University for Advanced Studies,
Tsukuba 305-0801, Japan}
}
\date{\normalsize November, 2000}
\maketitle
\thispagestyle{empty}

\begin{abstract}
\normalsize
We analyze how lower-dimensional bosonic D-branes further decay,
using the boundary string field theory.
Especially we find that the effective tachyon potential of the
lower-dimensional D-brane has the same profile as that of D25-brane.
\end{abstract}

\end{titlepage}

\section{Introduction}
Since the discovery of D-brane \cite{Pol}, our understanding of string
theory has been deepened.
Among other things it is now possible to discuss the open string
tachyon \cite{Sen1,Sen2,Sen3,Sen}.
It was conjectured that bosonic D-brane of any dimension can decay
into the closed string vacuum or lower-dimensional D-brane.
Moreover the vacuum energy of the bosonic D-brane is considered to
correspond to the tension of the D-brane.

Old days calculation \cite{KosSam} in the open string field theory
\cite{CSFT} has been renewed to discuss the tachyon condensation
\cite{SenZwi,Tay,MoeTay,HarKra,dMKJMT,MSZ,dMKR,Moe}.
Since in these cases all scalar quantities may acquire vacuum
expectation values, we can only analyze the tachyon condensation by
truncating the infinite levels of string excitations.
Some attempts for the exact manipulation are found in
\cite{universality,RZ,KP,HS}.

However, recently this difficulty has been overcome
\cite{GerSha,KMM1,GhoSen}.
Using another formulation called boundary string field theory (BSFT)
\cite{BSFT1,BSFT2,BSFT3,BSFT4,BSFT5} we have only to consider the
tachyon field in discussing the tachyon condensation.
This is because the general property of the renormalization group flow
ensures that the quadratic modes of the tachyon field decouple from
the other modes.
Exact analysis was performed in this formulation.
Both the situations that D25-brane decays into the closed string
vacuum and that D25-brane decays into lower-dimensional brane are
analyzed, and Sen's conjectures relating the vacuum energy and the
brane tension are confirmed exactly.
The derivative truncated effective action of the tachyon field
obtained from BSFT is found to agree with the toy model proposed in
\cite{Zwi,MZ1}.
Several related works are also found in
\cite{MZ2,Cor,Oku,KMM2,And,DD}.

In this paper we use this formulation to discuss the descent relation
of tachyon condensation.
We analyze, after D25-brane decays into lower-dimensional brane, how
it further decays.
Especially we find that the effective tachyon potential of
the lower-dimensional brane has the same profiles as that of
D25-brane.

In the next section, we review some results of BSFT for later
necessity. After the review we proceed in Sec.\ 3 to analyze the
descent relation using BSFT. Here our method may seem unusual from the
field theoretical viewpoint. Hence in Sec.\ 4, we restate the results
using the field theoretical analysis and furthermore justify our
method used in Sec.\ 3. We conclude in the final section.

\section{Boundary String Field Theory}
In this section, we shall review some results of BSFT
\cite{BSFT1,BSFT2,BSFT3,BSFT4,BSFT5} which is necessary for later
analysis.
The BSFT is constructed \cite{BSFT1} by identifying several contents
in the Batalin-Vilkovisky (BV) formalism with those in string field
theory.
Hence we shall first shortly recall the BV formalism.
The BV formalism is defined as follows.
Assume that a supermanifold $\calM$ with local coordinates $\lambda^i$
and fermionic closed 2-form $\omega_{ij}$ is given.
If the action $S$ satisfies the master equation $\{S,S\}=0$ with the
antibracket defined as
$\{A,B\}=A\overleftarrow{\p_i}\omega^{ij}\overrightarrow{\p_j}B$,
the action $S$ will necessarily have gauge symmetry.
The existence of the action $S$ is equivalent to the existence of a
vector field $V^i$ satisfying the nilpotency condition $V^2=0$ and the 
local integrability condition $d(i_V\omega)=0$.
The two quantities $S$ and $V^i$ are related via $i_V\omega=dS$.

To construct BSFT we have to identify the contents in the BV formalism
with those in string field theory.
We first identify the supermanifold $\calM$ as the space of open
string fields $\{\calO_i\}$ and the fermionic 2-form
$\omega_{ij}$ as the two point function of the string fields 
$\omega_{ij}=\langle\calO_i\calO_j\rangle$.
Here the string field $\calO$ has ghost number one and is related to
the boundary perturbation $S_{\p\Sigma}=\int{d\theta}\calV$ by the
relation $\calO=c\calV$.
Besides if the vector $V^i$ is identified as the BRST current vector,
the derivative of the action is given as
\bea
dS=\frac12\int{d\theta}{d\theta'}
\Bigl\langle d(c\calV)(\theta)
\{Q_{\rm BRST},c\calV\}(\theta')\Bigr\rangle_{\calV}.
\eea
It is possible to integrate this derivative as \cite{BSFT4}
\bea
S(\lambda)=Z(\lambda)
\biggl(1+\beta^i(\lambda)\frac{\p}{\p\lambda^i}\log Z(\lambda)\biggr),
\eea
with some beta functions $\beta^i(\lambda)$ and the partition function
$Z(\lambda)$ given as
\bea
Z(\lambda)=\int\calD X
\exp\Bigl(-S_\Sigma(X)-S_{\p\Sigma}(X,\lambda)\Bigr).
\eea

In the special case of the tachyon field $T$ in the quadratic
profile,
\bea
T=a+\sum_i\frac{u_i}{2\alpha'}X_i^2,
\label{quad}
\eea
the BSFT action is given as \cite{BSFT2}
\bea
S(a,u_i)=e^{-a}\prod_iZ_1(u_i)\biggl(1+a+\sum_iu_i
-\sum_iu_i\frac{\p}{\p u_i}\log Z_1(u_i)\biggr),
\label{BSFT}
\eea
where $Z_1(u)$ is defined as
\bea
Z_1(u)\equiv\sqrt{u}e^{\gamma u}\Gamma(u).
\eea
If we expand this action with respect to the coordinate $u$ up to the
next to leading order,
\bea
S(a,u_i)=\Bigl(14+a+\sum_iu_i+O(u^2)\Bigr)
e^{-a}\prod_i\frac{1}{\sqrt{u_i}}
\label{nexttolead}
\eea
we can rewrite the action in the field theoretical form:
\bea
S=T_{25}\int d^{26}x\Bigl(\alpha'e^{-T}(\vec\p_xT)^2+e^{-T}(T+1)\Bigr),
\eea
or
\bea
S=4e\cdot T_{25}\int d^{26}x
\biggl(\alpha'(\vec\p_x\phi)^2
-\frac14\phi^2\log\phi^2\biggr),
\label{field}
\eea
with the D25-brane tension $T_{25}$ defined as $1/(2\pi\alpha')^{13}$
in this case. (See \cite{KMM1,GhoSen}.)
Here we have normalized the kinetic term in the usual way via the
field redefinition $\log\phi^2=-(T+1)$.
The coordinates $x^i$ of the target space correspond to the zero modes
of the worldsheet fields $X^i$.

The exact BSFT action (\ref{BSFT}) is analyzed carefully in
\cite{KMM1}.
If we set all $u_i$ zero except only one direction $u_1=u$, we can
depict the profile of the action as in Fig.\ 1.
Here three stationary points are found and they correspond to the
perturbative vacuum of D25-brane ($a=0$, $u=0$), D24-brane
($a=\infty$, $u=\infty$) and the closed string vacuum where open
strings condense ($a=\infty$, $u=0$), respectively.
When D25-brane decays into D24-brane we have to take both $a$ and $u$
to infinity under the relation
\bea
a=-u+u\frac{\p}{\p u}\log Z_1(u),
\label{statcond}
\eea
which is the stationary condition of $a$.
\begin{figure}[hbt]
\begin{center}
\leavevmode
\epsfxsize=60mm
\put(100,15){$a$}
\put(170,80){$u$}
\put(-40,100){$S(a,u)$}
\epsfbox{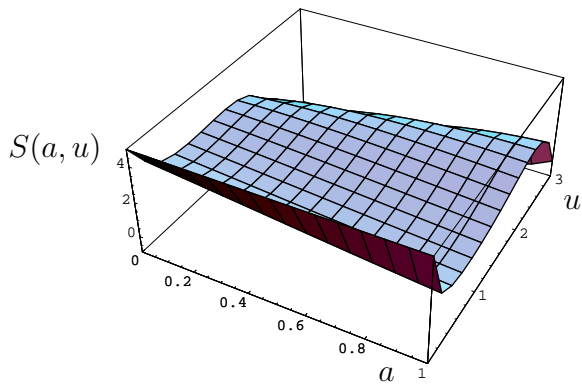}
\caption{The profile of the BSFT action $S(a,u)$. The D24-brane is
  expressed as the stationary point $a,u\to\infty$ along the ridge
  (\ref{statcond}).}
\end{center}
\end{figure}

\section{Descent Relation from Boundary String Field Theory}
Here we shall analyze how lower-dimensional D-brane further decays.
For simplicity we shall study the effective tachyon potential and
further decays of D24-brane.
Therefore we keep only two directions $T=a+(uX^2+vY^2)/2\alpha'$ for
consideration: one ($u_1=u$, $X_1=X$) is used when D25-brane decays
into D24-brane and the other ($u_2=v$, $X_2=Y$) is kept to discuss
further decays.
Hence the previous action $S(a,u_i)$ is rewritten as $S(a,u,v)$.
The generalizations to other decays are straightforward.

First note that after D25-brane decays into D24-brane both the
coordinates $a$ and $u$ go to infinity.
However, if we would like to discuss further decay, it is necessary to
identify the coordinate that plays the role of the tachyon zero mode
of D24-brane, just like the coordinate $a$ in the case of D25-brane.
There must be some choices.
But as we shall discuss later, it should be natural to consider the
section of a large constant $u=\ust$ and use again the coordinate $a$
for the tachyon zero mode with the origin shifted by $\ast$.

More precisely, we shall evaluate $S(\ast+a,\ust,v)$ in the limit
$\ast,\ust\to\infty$. The double limit is taken under the relation
(\ref{statcond}) for $(\ast,\ust)$ since we consider the decay of
D24-brane here.
The calculation is straightforward if we use the asymptotic forms
around $u\sim\infty$:
\bea
\log Z_1(u)&\hspace{-2mm}\sim
&\hspace{-2mm}u\log u-u+\gamma u+\log\sqrt{2\pi},\nn\\
u\frac{\p}{\p u}\log Z_1(u)&\hspace{-2mm}\sim
&\hspace{-2mm}u\log u+\gamma u.
\eea
After a little algebra we find the following result:
\bea
\lim_{\ast,\ust\to\infty}S(\ast+a,\ust,v)=\sqrt{2\pi}e^{-a}Z_1(v)
\biggl(1+a+v-v\frac{\p}{\p v}\log Z_1(v)\biggl).
\label{lower}
\eea
Amazingly, this has exactly the same form as the action of the tachyon
field on D25-brane except that the dimension is lowered by one and the
factor $\sqrt{2\pi}$ appears.
Hence, here we have found heuristically that the profile of the
tachyon potential of D24-brane is the same as that of D25-brane.
Namely, the tachyon potential has self-similarity.
This result makes it possible to discuss further decays of D24-brane
in the same way as that of D25-brane.

We shall make a few comments here.

First our result here seems natural.
D-brane of any dimension can decay into lower-dimensional D-brane or
the closed string vacuum.
Hence it is natural to expect that D-brane of any dimension should
all have the same structure in the tachyon mode.

Secondly, as we have mentioned before our analysis, there are some
choices to identify the tachyon zero mode of D24-brane.
For example, though here we have fixed the coordinate $u$ to be a
large constant $\ust$, we can alternatively fix the coordinate
$a=\ast$ and try to regard the coordinate $u$ as the tachyon zero
mode.
This time we have to evaluate the action $S(\ast,\ust-u,v)$ in the
limit $\ast,\ust\to\infty$.
Although in this case we also find the self-similarity, it is not the
coordinate $u$ but
\bea
\ae\equiv\ust\log\ust-(\ust-u)\log(\ust-u)+(\gamma-1)u
\label{aeff}
\eea
that plays the role of the tachyon zero mode.
However, as we shall discuss in the next section, it is most natural
to identify the coordinate $a$ as the tachyon zero mode as in
(\ref{lower}).

\section{Field Theoretical Analysis}
In the previous section we have analyzed how the lower-dimensional
D-brane further decays using the exact BSFT action.
We have found that the profile of the tachyon potential is unchanged
under the descent relation.
However, what we have done is somewhat unusual from the field
theoretical viewpoint.
In the present section we would like to complement the previous
analysis using the field theoretical action (\ref{field})
\cite{Zwi,MZ1} obtained by truncating the higher derivatives.
In this method we can also obtain the same results.
Besides, it would shed light on our previous question:
what coordinate plays the role of the tachyon zero mode of the
lower-dimensional D-brane.
Therefore, our analysis here justifies the previous interpretation of
$S(\ast+a,\ust,v)$ as the action of the tachyon field on the
lower-dimensional D-brane.

The derivative truncated action (\ref{field}) is written as
\bea
S=4e\cdot T_{25}\int d^{25}ydx
\biggl(\alpha'(\vec\p_y\phi)^2+\alpha'(\p_x\phi)^2
-\frac14\phi^2\log\phi^2\biggr).
\label{trun}
\eea
As in the previous section here we also consider the effective action
of the tachyon field on D24-brane for simplicity.
The target space coordinate $x$ denotes the transverse direction of
the D24-brane and $\vec y$ the longitudinal directions.
The D24-brane is expressed as a classical lump solution
$\bar\phi(x)$.
To see the fluctuation on the D24-brane we have to expand the tachyon
field into infinite modes around this solution as
\bea
&&\phi(x,\vec y)=\bar\phi(x)+\tilde\phi(x,\vec y),\nn\\
&&\tilde\phi(x,\vec y)=\sum_n\xi_n(\vec y)\psi_n(x).
\label{fluc}
\eea

If we want to obtain the effective action for the lowest mode of the
D24-brane, we have to put the expansion (\ref{fluc}) with the
classical solution $\bar\phi(x)$ and the fluctuating modes
$\tilde\phi(x,\vec y)$ into the original derivative truncated action
(\ref{trun}) and integrate over the coordinate $x$.
As shown in \cite{MZ1}, when we calculate the effective action for the 
lowest mode, the effect of the higher excited modes does not come in.
It was also found that both the profiles of the lump solution
$\bar\phi(x)$ and the lowest mode of D24-brane $\psi_0(x)$ are the
gaussian functions:
\bea
\bar\phi(x)=\psi_0(x)=\exp\biggl(-\frac{1}{8\alpha'}x^2\biggr).
\label{gauss}
\eea
After a straightforward calculation we find the derivative truncated
action (\ref{trun}) becomes
\bea
S=4e\cdot T_{25}\sqrt{4\pi\alpha'}\int d^{25}y
\biggl(\alpha'\Bigl(\vec \p_y\Bigl(1+\xi_0(\vec y)\Bigr)\Bigr)^2
-\frac14\Bigl(1+\xi_0(\vec y)\Bigr)^2
\log\frac{\Bigl(1+\xi_0(\vec y)\Bigr)^2}{e}\biggr).
\eea
If we rewrite the lowest mode $\xi_0(\vec y)$ into a new field
$\phieff(\vec y)$ as
\bea
1+\xi_0(\vec y)\equiv\sqrt{e}\phieff(\vec y)
\label{fieldredef}
\eea
we find that the derivative truncated action is unchanged except that
the dimension of the field theory is lowered by one and the tension
becomes $T_{25}2\pi\sqrt{\alpha'}\cdot(e/\sqrt{\pi})$ which is
$T_{24}$ in the interpretation of \cite{MZ1,KMM1}.

Here we have repeated the analysis of the effective action of
D24-brane using the field theoretical method.
We have found that the result is perfectly consistent with our
expectation and the result of the previous section.

Let us return back to the question raised in the previous section:
why we have identified the coordinate $a$ as the tachyon zero mode
with the origin shifted.
To answer this question from the field theoretical analysis, first let
us see how the manipulation in this section relates to the
manipulation in the previous section.

In the analysis of BSFT in the previous section, we have first assumed
the quadratic profile for the tachyon field (\ref{quad}) and obtained
directly the integrated expression (\ref{BSFT}).
In obtaining the action of the tachyon field on D24-brane, we have
restricted ourselves in the section of a large constant $u=\ust$ and
considered $S(\ast+a,\ust,v)$.

On the other hand, in obtaining the effective tachyon action using the
field theoretical analysis in the present section, we have rewritten
the field $\phi(x,\vec y)$ as
$\exp(-x^2/8\alpha')\times(1+\xi_0(\vec y))$ and integrated over the
coordinate $x$.
In other words, since both the classical lump solution $\bar\phi(x)$
and the lowest fluctuating mode $\psi_0(x)$ have the same quadratic
exponent with fixed coefficient $1/8\alpha'$, it is possible to
factorize the field $\phi$ as the product of the classical solution
$\bar\phi$ in the direction $x$ and the fluctuating mode
$\phi_{\rm eff}$ in the direction $\vec y$ as
\bea
\phi=\bar\phi\left[\ast,\ust,0\right]
\times\sqrt{e}\phi_{\rm eff}\left[a,0,v\right].
\label{factor}
\eea
Here we have defined the notation $\phi\left[a,u,v\right]$ (and their
cousins $\bar\phi$ and $\phi_{\rm eff}$) as
\bea
\phi\left[a,u,v\right]\equiv\exp\biggl\{-\frac12
\biggl(a+\frac{u}{2\alpha'}x^2+\frac{v}{2\alpha'}\vec y^2+1\biggr)
\biggr\},
\eea
using the relation $\log\phi^2=-(T+1)$ and set $\ast=-1$ and
$\ust=1/2$ for the derivative truncated action (\ref{trun}).
Note that the right-hand-side of (\ref{factor}) is equal to
$\phi\left[\ast+a,\ust,v\right]$.
Hence, what we have done in this section is exactly to consider the
action $S(\ast+a,\ust,v)$ in the section of $\ust=1/2$ with the origin
of $a$ shifted by $\ast=-1$ from the viewpoint of the integrated form
(\ref{nexttolead}).
Here the coordinate $a$ again corresponds to the tachyon zero mode on
D24-brane.
This is nothing but what we did in the previous section.
In this way we have related the manipulation in this section to that
in the previous section and justified the method of Sec.\ 3.

However, it might be dangerous to generalize the above arguments
naively to the exact BSFT action.
The exact BSFT action (\ref{BSFT}) is written not as the field theory
but in the integrated form.
Hence, though we know that the classical solution should be quadratic,
it is difficult in principle to discuss the fluctuating modes around
the classical solution and see whether the effect of the higher
fluctuating modes in (\ref{fluc}) comes in.
Moreover, if we want to relate the two manipulations, the classical
lump solution $\bar\phi(x)$ and the lowest fluctuating mode
$\psi_0(x)$ should have the same gaussian profiles.
Here even in the exact case of the previous section, we have
optimistically assumed these properties and analyzed
$S(\ast+a,\ust,v)$ in the same way as the derivative truncated case.

\section{Conclusion}
In this paper we have analyzed the descent relation of the tachyon
condensation.
Especially we have found that the profile of the tachyon potential of
lower-dimensional D-brane is the same as that of D25-brane.
Namely, the tachyon potential has self-similarity.

In discussing the effective action of the tachyon field on the
lower-dimensional D-brane using BSFT in Sec.\ 3, we had to know how to
deal with the tachyon zero mode.
Here we made a non-trivial suggestion that we should shift the origin
of the tachyon zero mode and consider the BSFT action
$S(\ast+a,\ust,v)$ in the limit $\ast,\ust\to\infty$.

Furthermore, we repeated the analysis in the field theoretical model
in Sec.\ 4.
We found that what we did in the field theoretical model is
essentially to shift the origin of the tachyon zero mode and consider
the derivative truncated action $S(\ast+a,\ust,v)$ with $\ast=-1$ and
$\ust=1/2$.
This is nothing but the method used in Sec.\ 3.
Hence, we also justified the above suggestion, using the more
familiar field theoretical analysis.

Finally let us give a comment on the structure of the closed string
vacuum.
When we discuss the situation that D25-brane decays into the closed
string vacuum, we set all $u_i$ zero and consider the coordinate $a$
becomes infinity.
However, if we discuss the decay of D24-brane into the closed string
vacuum, first we have to consider $a\to\infty$ with a large constant
$u=\ust$ and send $\ust\to\infty$ afterwards.
Hence, our closed string vacuum in this case seems to be obtained in
the limit $a,u\to\infty$ and different from the original closed string
vacuum ($a=\infty$, $u=0$).
This is a very delicate problem that relates to how we take the
double limit.
To clarify the vacuum structure at $a=\infty$ should be an interesting 
direction.

\noindent
{\large\bf Acknowledgment}

We would like to thank Y.\ Michishita, S.\ Shinohara and especially
H.\ Hata for valuable discussions and comments.
This work is supported in part by Grant-in-Aid for Scientific Research
from Ministry of Education, Science, Sports and Culture of Japan
(\#04633).
One of the authors (S.\ M.) is supported in part by the Japan Society
for the Promotion of Science under the Predoctoral Research Program.

\newcommand{\J}[4]{{\sl #1} {\bf #2} (#3) #4}
\newcommand{\andJ}[3]{{\bf #1} (#2) #3}
\newcommand{\AP}{Ann.\ Phys.\ (N.Y.)}
\newcommand{\MPL}{Mod.\ Phys.\ Lett.}
\newcommand{\NP}{Nucl.\ Phys.}
\newcommand{\PL}{Phys.\ Lett.}
\newcommand{\PR}{Phys.\ Rev.}
\newcommand{\PRL}{Phys.\ Rev.\ Lett.}
\newcommand{\ATMP}{Adv.\ Theor.\ Math.\ Phys.}

\end{document}